\documentclass[showpacs,preprint,aps]{revtex4}
\usepackage{graphics}
\usepackage{url}
\usepackage{rotating}
\usepackage{graphicx}
\usepackage{bm}
\usepackage{epsfig}
\usepackage{amsmath}
\usepackage{subfigure}

\def\be{\begin{equation}}
\def\lan{\left\langle}
\def\ran{\right\rangle}
\def\ee{\end{equation}}
\def\barr{\begin{array}}
\def\earr{\end{array}}

\def\nn8{\\}
\def\l{\left}
\def\r{\right}
\def\dis{\displaystyle}
\def\ed{\end{document}}

\begin{document}

\title{Deformed shell model results for two neutrino positron double beta
decay of $^{84}$Sr}

\author{R. Sahu$^1$, and V.K.B. Kota$^{2,3}$}

\affiliation{$^1$Physics Department, Berhampur University,  Berhampur 760
007, Orissa, India \\ $^2$Physical Research Laboratory, Ahmedabad 380 009,  
India \\ $^3$Department of Physics, Laurentian  University, Sudbury, ON P3E
2C6, Canada}

\begin{abstract}

Half-lives $T_{1/2}^{2\nu}$ for two neutrino positron double beta decay
modes $\beta^{+}$EC/ECEC are calculated for $^{84}$Sr, a nucleus of current
experimental interest, within the framework of the deformed shell model
based on Hartree-Fock states employing a modified Kuo interaction in
($^{2}p_{3/2}$, $^{1}f_{5/2}$, $^{2}p_{1/2}$, $^{1}g_{9/2}$) space. For a
reasonable description of the spectra of $^{84}$Sr and $^{84}$Kr and to
generate allowed GT strengths, the  single particle energies of the proton
and neutron $^1g_{9/2}$ orbitals, relative to the $^{2}p_{3/2}$ orbital
energy, are chosen to be ($3.5$ MeV, $1.5$ MeV) for both $^{84}$Sr and
$^{84}$Rb and ($1.5$ MeV, $1.5$ MeV) for $^{84}$Kr. With this, the
calculated half-lives for the $\beta^+$EC and ECEC modes are $\sim
10^{26}$yr and $\sim 4 \times 10^{24}$yr respectively. 

\end{abstract}



\maketitle
\section{Introduction}

Double beta decay (DBD) is a rare weak interaction process in which two
identical nucleons inside the nucleus undergo decay either  with the emission
of two neutrinos or without any neutrinos. The two neutrino double beta decay
($2\nu \beta ^{-}\beta ^{-})$ which was first predicted long back by Meyer
\cite{y1} has been observed  experimentally in more than 10 nuclei
\cite{y2,y3}.  In contrast, the positron decay modes, i.e. $2\nu\; \beta
^{+}\beta^{+}/\beta^{+}$EC/ECEC decay modes (hereafter, all these three
combined is called $2\nu\;e^+$DBD) are not yet observed experimentally
(exception  being $^{130}$Ba decay  derived from geochemical methods
\cite{Me-01}).  However, in the last few years there are serious attempts,
using direct counting methods,   to measure half-lives for $2\nu\;e^+$DBD
modes in many nuclei ranging from $^{64}$Zn to $^{132}$Ba
\cite{zn64,se74,kr78,cd106,ba130}. In the mass A=80 region, the  candidate
nuclei for $2\nu\;e^+$DBD  are $^{78}$Kr, $^{74}$Se and $^{84}$Sr. For
$^{78}$Kr and $^{74}$Se there are already some experimental efforts and
following this,  recently \cite{y5-kr,y5-se} we have carried out calculations
for these two nuclei using the so called deformed shell model (DSM). In
addition we had previously  attempted to study $2\nu \beta ^{-}\beta ^{-}$
nuclear matrix elements for $^{76}$Ge $\rightarrow$ $^{76}$Se using DSM 
\cite{app2}. There was also an attempt for $^{82}$Se $\rightarrow$ $^{82}$Kr
and it is seen that \cite{Sa-10} DSM produces nuclear matrix elements within
a factor of 2 the QRPA results given in \cite{Bo-00}.

Over the years, the DSM model based on Hartree-Fock states has been used to
study with considerable  success: (i) spectroscopic properties, such as band
structures, shapes, nature of band crossings, electromagnetic transition
probabilities and so on in A=64-80 nuclei
\cite{sahu1,sahu2,sahu3,br8082,npa96,sahu4}; (ii) $T=1$ and  $T=0$ bands in
N=Z odd-odd nuclei and $T=1/2$ bands in odd-A nuclei by including isospin
projection \cite{sk1,sk2,msk};  (iii)  transition matrix elements for $\mu-e$
conversion in $^{72}$Ge  \cite{app1} and in the analysis of data for
inelastic scattering of electrons from $fp$-shell nuclei \cite{app3};  (iv)
$\beta$-decay half lives, GT distributions and electron capture rates in N
$\sim$ Z nuclei in A=60-80 region \cite{y5-kr} and so on. It is seen that 
the predictions of DSM for  $2\nu\;e^+$DBD half-lives for $^{78}$Kr are close
to those of QRPA  and PHFB models. For $^{74}$Se there are no other model
predictions except those of DSM. See \cite{y5-kr,y5-se} for details. 

A group led by H.J. Kim at Korean KPNU facility has initiated DBD
experiments for positron double beta decay (both $2\nu$ and $0\nu$)  in
$^{84}$Sr using  SrCl$_2$ crystals \cite{kim1,kim2}. Prompted by these
efforts,   we have carried out DSM  calculations, extending the studies in
\cite{y5-kr,y5-se} further, for $2\nu\;e^+$DBD half-lives for $^{84}$Sr
nucleus and the  results are reported in this paper. It should be pointed
out that there are no theoretical results available in literature for this
nucleus. Just as with $^{78}$Kr and $^{74}$Se, we have undertaken the
$2\nu\;e^+$DBD study of $^{84}$Sr as it was already  shown that DSM model
describes very well \cite{sahu} the low-lying bands in this nucleus. Also
as we shall discuss ahead, the proton and neutron single particle energies
(spe)  of the $^1g_{9/2}$ orbital play an important role in describing the
spectroscopic results of  $^{84}$Kr and in determining the $^{84}$Sr
$\rightarrow$ $^{84}$Kr  $2\nu\;e^+$DBD half lives. Now we will give a
preview.

In Section II given is, for completeness, a brief discussion of DSM model
and  the formalism for calculating $2\nu\;e^+$DBD half-lives. Spectroscopic
results  for $^{84}$Sr and $^{84}$Kr are given in Section III. In Section
IV, DSM results for $2\nu\;e^+$DBD half  lives are discussed and also given
here are some concluding remarks.

\section{$2\nu\;e^+$DBD half-lives and deformed shell model}

Half-life for the $2\nu \;e^{+}$DBD decay modes for the $0_I^{+}  \rightarrow
0_F^{+}$ transitions, is given by \cite{doi92},
\be
\l[ T_{1/2}^{2\nu}(k)\r] ^{-1} = G_{2\nu }(k) \;\l|M_{2\nu }\r|^2
\ee
where $k$ denotes the modes $\beta ^{+}\beta ^{+}$, $\beta ^{+}$EC and ECEC. 
The kinematical factors $G_{2\nu}(k) $ are  independent of nuclear structure
and they can be calculated with good accuracy \cite{doi92,boe92}. On the
other hand, the nuclear transition matrix elements (NTME) $M_{2\nu}$ are
nuclear model dependent and they are given by \cite{doi92,Hir-94}, 
\be
M_{2\nu}=\;\sum\limits_N\;\dis\frac{\lan 0_F^+ \mid\mid \mathbf{\sigma}
\tau^{-} \mid\mid 1_N^+ \ran \lan 1_N^+ \mid\mid \mathbf{\sigma}\tau
^{-} \mid\mid 0_I^+ \ran}{\l[ E_0 + E_N - E_I\r]}
\label{m2n}
\ee
where $\l| 0_{I}^{+}\ran$, $\l| 0_{F}^{+}\ran$ and  $\l| 1_{N}^{+}\ran$ are
the initial, final and virtual intermediate states respectively and
$E_{N}(E_{I})$ is the energy of the intermediate (initial) nucleus.
Similarly, $E_{0}=\frac{1}{2}W_{0}$ where $W_{0}$ is the total energy
released for different $2\nu \;e^{+}$DBD modes. As given  in
\cite{Hir-94,Shuk1,shuk2}, $W_{0}(\beta ^{+}\beta ^{+})=Q_{\beta ^{+}\beta
^{+}}+2m_{e}$, $W_{0}(\beta ^{+}\mbox{EC}) = Q_{\beta ^{+}\mbox{EC}} + e_{b}$
and $W_{0}(\mbox{ECEC}) = Q_{\mbox{ECEC} } - 2m_{e} + e_{b1} + e_{b2}$. The
Q-values for different $2\nu \;e^{+}$DBD modes are \cite{Hir-94}: $Q_{\beta
^{+}\beta ^{+}} = M(A,Z) - M(A,Z-2) - 4m_{e}$, $Q_{\beta ^{+}EC} = M(A,Z) - 
M(A,Z-2) - 2m_{e}$ and $Q_{ECEC} = M(A,Z) - M(A,Z-2)$ . Here $M$ denotes the
neutral atomic mass (available for example from the  tabulations in
\cite{Aud03}) and $e_{b}$ is the binding energy of the captured atomic
electron.  Energies in the denominator in Eq. (\ref{m2n}) are taken in the
units of electron mass. With the atomic mass difference being $1787$ keV for
$^{84}$Sr decay, it should be clear that the $\beta^+ \beta^+$ DBD mode is
forbidden (as the $Q$-value will be negative) for $^{84}$Sr. For $\beta^+$EC 
and ECEC modes, one and two K shell  electron capture respectively are
considered. The binding energy for the K shell electrons is taken from
\cite{be}.

In DSM, for a given nucleus, starting with a model space consisting of a
given set of single particle (sp) orbitals and effective two-body
Hamiltonian, the lowest prolate and oblate intrinsic states are obtained by
solving the Hartree-Fock (HF) single particle equation self-consistently.
Excited intrinsic configurations are obtained by making particle-hole
excitations over the lowest intrinsic state. These intrinsic states will not
have good angular momentum and hence good angular momentum states are
obtained by angular momentum projection from these intrinsic states. The
normalized states of good angular momentum projected from the intrinsic state
$\chi_K(\eta)$ can be written in the form
\be
\psi^J_{MK}(\eta) = \frac{2J+1}{8\pi^2\sqrt{N_{JK}}}\int d\Omega D^{J^*}_
{MK}(\Omega)R(\Omega)\l| \chi_K(\eta) \ran
\label{eq.proj}
\ee
where $N_{JK}$ is the normalization constant given by
\be
N_{JK} =\frac{2J+1}{2} \int^\pi_0 d\beta \sin \beta d^J_{KK}(\beta)
\lan \chi_K(\eta)|e^{-i\beta J_y}|\chi_K(\eta)\ran \;.
\ee
In Eq. (\ref{eq.proj}) $\Omega$ represents the Euler angles ($\alpha$,
$\beta$, $\gamma$), $R(\Omega)$ which is equal to exp($-i\alpha
J_z$)exp($-i\beta J_y$)exp( $-i\gamma J_z$) represents the general rotation
operator. In general the projected states with same $J$ but coming from
different intrinsic states will not be orthogonal to each other. Hence they
are orthonormalized and then band mixing calculations are performed.  DSM is
well established to be a successful model for transitional nuclei (with
A=64-80) when sufficiently large number of intrinsic states are included in
the band mixing calculations. Performing DSM calculations for the parent,
daughter and the intermediate odd-odd nucleus (here we need only the $1^+$
states) and then using the DSM wavefunctions, the $\sigma\tau^-$ matrix
elements in Eq. (2) are calculated. For further details of DSM see
\cite{y5-kr}. 

Let us add that the recently introduced projected configuration interaction
(PCI) model of Horoi et al \cite{pci1,pci2} and the projected shell model
(PSM) of Sun et al \cite{psm1,psm2,psm3} are quite similar to DSM (PCI may
be better for quasi-spherical nuclei and PSM includes BCS correlations from
the beginning).

\section{Spectroscopic results for $^{84}$Sr and $^{84}$Kr}

\subsection{$^{84}$Sr}

In the DSM calculations  for the structure of the nuclei $^{84}$Sr,
$^{84}$Rb and $^{84}$Kr and for $^{84}$Sr $2\nu \;e^{+}$DBD half-lives
presented  in section IV,  we have used a modified Kuo effective interaction
\cite{int} in the ($^{2}p_{3/2}$,  $^{1}f_{5/2}$, $^{2}p_{1/2}$,
$^{1}g_{9/2}$) space with $^{56}$Ni as the inert core. The spe of these
orbitals are taken as 0.0, 0.78, 1.08 and 3.5 MeV respectively in a first
set of calculations and in the next set of calculations,  the $^{1}g_{9/2}$
energy for the protons and/or neutrons is changed to 1.5 MeV  for reasons
discussed in detail ahead. DSM with modified Kuo effective interaction has
been quite successfully used by us in describing many important features of
nuclei in A $\sim $ 60-80 region. In particular, shape coexistence in
spectra, observed $B(E2)$ values, identical bands, band crossings, and
deformations in   $^{76,78,80,82,84}$Sr isotopes are well described by DSM
\cite{sahu,sahu3}.  First  we will describe briefly the earlier
spectroscopic results for $^{84}$Sr. 

Starting with the modified Kuo interaction and $^{2}p_{3/2}$,  $^{1}f_{5/2}$,
$^{2}p_{1/2}$, $^{1}g_{9/2}$ spe (both for protons and neutrons) to be  0.0,
0.78, 1.08 and 3.5 MeV respectively, DSM calculations have been carried out
for $^{84}$Sr in \cite{sahu}. Here, first axially symmetric HF  calculations
are performed and the lowest prolate HF intrinsic state is obtained. The
reason for neglecting the oblate states has been discussed in detail in
\cite{npa96}. The lowest HF sp spectrum for $^{84}$Sr is shown in Fig. 1a;
note that the HF sp states with a given $k$ values will be doubly degenerate.
By particle-hole excitations from the  lowest intrinsic state shown in Fig.
1a, excited intrinsic states are generated. In the band mixing  calculations
a limited number of five configurations are taken and they are: (i) ground
$K^\pi=0^+$ intrinsic state; (ii) two excited $K^\pi=0^+$ intrinsic states
obtained by exciting two or four protons into the $k$ states arising out of
the $^{1}g_{9/2}$ orbit;  (iii) a $K^\pi=8^+$ intrinsic state obtained by
exciting  two valence neutrons into $k=7/2^+$ and $9/2^+$ states; (iv) a
$K^\pi=2^+$ intrinsic state obtained by exciting a valence proton from
$k=5/2^-$ into $1/2^-$ state. As discussed in detail in \cite{sahu}, the band
mixing calculations reproduce the observed data quite well. Figure 2 shows
the calculated and experimental spectra.
The DSM results in Fig. 2 are slightly different from those in Fig. 2a  of
\cite{sahu} where the so called tagged HF calculation is done (in the present
calculations, just as in \cite{y5-kr,y5-se}, no tagging has been done). The
results for $10^+$ and $12^+$ levels are in better agreement with data
compared to the earlier results in \cite{sahu}. The two lowest excited $8^+$
bands in Fig. 2 are established by experiments \cite{dew-82,nndc}  to be two
neutron and two proton aligned bands respectively and DSM reproduces this
structure. Similarly, the observed B(E2) values and also the excited $2^+$
band are well described. All these confirm that DSM gives good spectroscopic
results for $^{84}$Sr. 

In the  $2\nu \;e^{+}$DBD decay, the intermediate nucleus is $^{84}$Rb. Fig.
1b gives the HF spectrum for $^{84}$Rb (for completeness we also show the HF
spectrum for $^{84}$Kr in Fig. 1c).  As we can see from Fig. 1, in the DBD
process, in the lowest order,  the valence proton from the $5/2^-$ orbit in
$^{84}$Sr will go to the unoccupied neutron $7/2^+$ orbit in $^{84}$Rb. This
gives a negative parity state for $^{84}$Rb and this is not allowed by GT.
Hence an intrinsic state with a proton in $5/2^+$ in $^{84}$Sr is needed to
generate  allowed GT matrix elements. However the intensity of this
configuration will be small in the $^{84}$Sr ground state. Thus the GT
matrix elements will be small for $^{84}$Sr to $^{84}$Rb. Therefore we need
to reduce the neutron $^1g_{9/2}$ spe for these  two nuclei so that the
negative parity neutron sp states will be near the Fermi surface. As a
result we will be generating many low-lying $K^\pi = 1^+$ levels in
$^{84}$Rb and there will be large GT matrix elements for $^{84}$Sr to
$^{84}$Rb (similarly also for $^{84}$Rb to $^{84}$Kr). Following this, we
have considered proton $^1g_{9/2}$ spe $\epsilon^{(p)}(^1g_{9/2}) =3.5$ MeV
and neutron $^1g_{9/2}$ spe $\epsilon^{(n)}(^1g_{9/2}) =1.5$ MeV.  The HF sp
spectrum for this choice is shown in Fig. 3a. Starting with the lowest
configuration shown in Fig. 3a,  five more excited configurations are
considered in the band mixing calculations  as before for $^{84}$Sr. Fig. 2
shows that practically there is no effect of the change in
$\epsilon^{(n)}(^1g_{9/2})$  on the spectrum (also wavefunctions and hence
the results in \cite{sahu} are well preserved). The reason being that there
is essentially no change in the proton and neutron occupancies. In Fig. 2,
deviations of the results of the earlier calculation from the present
calculation are shown in the parentheses with an asterisk. Deviations below
50 keV are not shown. 

In conclusion, for $^{84}$Sr and $^{84}$Rb nuclei, for DSM calculations we
employ the spe for  $^{2}p_{3/2}$,  $^{1}f_{5/2}$ and $^{2}p_{1/2}$ orbitals
(same spe for both protons and neutrons) to be 0.0, 0.78, 1.08 MeV
respectively while $[\epsilon^{(p)}(^1g_{9/2}),\epsilon^{(n)}(^1g_{9/2})] 
=[3.5, 1.5]$ MeV.  Now we will consider $^{84}$Kr where there were no
earlier DSM results.

\subsection{$^{84}$Kr}

Employing the same set of spe as chosen for $^{84}$Sr and $^{84}$Rb, the
lowest prolate HF intrinsic state is obtained for $^{84}$Kr; note that as
above  $\epsilon^{(p)}(^1g_{9/2}) =3.5$ MeV and $\epsilon^{(n)}(^1g_{9/2})
=1.5$ MeV. Then the HF sp spectrum in Fig. 1c remains unchanged except that
the HF neutron $1g_{9/2}$ $k$-states move down in energy  by $2$ MeV. By
particle-hole excitations from the lowest intrinsic state, excited intrinsic
states are generated for $^{84}$Kr and  band mixing  calculations are carried
out using lowest 15 configurations. The observed lowest $6^+$, $8^+$ and
$10^+$ states are two-proton aligned states and similarly the $8^+_2$ and
$10_2^+$  are two-neutron aligned bands. It is seen that the calculated
two-proton aligned band is high (by more than 1 MeV)  compared to experiment.
Also the $6^+$ level produces weak $B(E2)$ for $6^+_1 \rightarrow 4^+_1$; the
calculated value is 0.2 W.u. while 6.9 W.u. is the data value (see Table 1).
In order to bring the two-proton aligned band head to be close to data, it is
necessary to lower the $\epsilon^{(p)}(^1g_{9/2})$ energy to $1.5$ MeV in
addition to using  $\epsilon^{(n)}(^1g_{9/2}) =1.5$ MeV. The HF sp spectrum
with  $[\epsilon^{(p)}(^1g_{9/2}), \epsilon^{(n)}(^1g_{9/2})]= [1.5,1.5]$ MeV
is shown in Fig. 3c and the energy spectrum is shown in Fig. 4. In the DSM 
calculations, starting with the lowest configuration shown in Fig. 3c, 15
excited configurations are considered for band mixing. It is seen from Fig. 4
that the  two-proton aligned band is in reasonable agreement with experiment.
From Fig. 4 and also the B(E2) results in Table 1, we conclude that DSM gives
a reasonably good description of the spectroscopy of $^{84}$Kr with the
choice $[\epsilon^{(p)}(^1g_{9/2}), \epsilon^{(n)}(^1g_{9/2})]= [1.5,1.5]$
MeV. Let us add that the recent data on occupancies in $^{76}$Ge and
$^{76}$Se do point out that  \cite{Sch-08,Kay-09} $1g_{9/2}$ is probably much
closer to the $fp$ shell than anticipated in the past, thus justifying the
lowering of the energies of proton and neutron $1g_{9/2}$  orbitals. It is
also pluasible that part of the reason for these lower energies may be due to
the neglect of excitations into $1g_{7/2}$ and $2d_{5/2}$ orbitals.

In conclusion, for $^{84}$Kr for DSM calculations we employ the spe for 
$^{2}p_{3/2}$,  $^{1}f_{5/2}$ and $^{2}p_{1/2}$ orbitals (same spe for both
protons and neutrons) to be 0.0, 0.78, 1.08 MeV respectively while
$[\epsilon^{(p)}(^1g_{9/2}),\epsilon^{(n)}(^1g_{9/2})]  =[1.5, 1.5]$ MeV . 

\section{Results for $2\nu\;e^+$DBD half lives and conclusions}

Using the DSM wavefunctions, $2\nu $ $\beta ^{+}$EC and ECEC half-lives for
$^{84}$Se $\rightarrow $ $^{84}$Kr transitions are calculated and the
results are shown in Table 2. In the DSM calculations for the intermediate
nucleus $^{84}$Rb,  24 $K^\pi=1^+$ bands are mixed and they span 3 MeV from
the lowest $K^\pi=1^+$ band. The resulting  24 $1^+$ states are employed in
Eq. (2) to calculate the GT matrix elements. We have verified that the
inclusion of additional $1^+$ states do not change significantly the
half-life results. Further, the $0^+$ ground state of $^{84}$Sr is generated
by mixing 30 intrinsic states.  Similarly for generating the $0^+$ ground
state of $^{84}$Kr 11 intrinsic states are used in the DSM calculations.  
The Kinematical factors $G_{2\nu}$  are taken from \cite{boe92}. It is seen
from Table 2 that the predicted half-lives are   $\sim 10^{25}$yr and this
should be of interest for  future experiments. Instead of the choice
$[\epsilon^{(p)}(^1g_{9/2}), \epsilon^{(n)}(^1g_{9/2})]  =[1.5, 1.5]$ MeV,
if we use $(3.5, 1.5)$MeV for generating the ground state $0^+$ of $^{84}$Kr
just as the spe used for  ($^{84}$Sr, $^{84}$Rb), the half-lives reduce by a
factor $\sim2$ as shown in the brackets in Table 2.  Let us add that the
results in Table 2 are the first nuclear structure results for $2\nu $
$\beta ^{+}$EC and ECEC half-lives for $^{84}$Se $\rightarrow $ $^{84}$Kr. 

In conclusion, prompted by recent experimental interest  in $^{84}$Sr
$2\nu\;e^+$DBD and the results reported in \cite{y5-kr,y5-se} for $^{78}$Kr
and $^{74}$Se using the DSM model, we have carried out spectroscopic
calculations and then the DBD half-lives calculations for $^{84}$Sr
$\rightarrow$ $^{84}$Kr. We have shown that the DSM model gives reasonably
good spectroscopy for both $^{84}$Sr and $^{84}$Kr by adjusting the spe of
proton and neutron $1g_{9/2}$ orbitals. Proceeding further we have presented
the DSM results for half-lives for $\beta^+$EC and ECEC modes. These results
should be a good guide for the experiments initiated by Kim et al
\cite{kim1,kim2} at KPNU, Korea. Finally, formulation within DSM, for
calculating $0\nu$DBD and $0\nu\;e^+$DBD half lives is being developed and
the results for $^{84}$Sr $0\nu\;e^+$DBD half lives will be reported in
future.

\acknowledgments

Thanks are due to H.J. Kim for his interest in the results in the paper and
for correspondence. Thanks are also due to P.C. Srivastava for help in
preparing the figures. RS is thankful to DST (India) for financial support.

\newpage
\begin{table}
\begin{center}
\caption{B(E2) values for $^{84}$Kr in Weisskopf units (W.u.). DSM results
are compared with experimental results (called EXPT in the table) from
\cite{nndc}. The effective charges employed are $e_p=1.6e$ and $e_n=1.0e$
just as in the previous paper on $^{84}$Sr \cite{sahu}.}
\begin{tabular}{lccc}
\hline 
Transition & DSM & EXPT \\
\hline
$2^+_1 \rightarrow 0^+_1$ & 18.4 & 12 \\
$4^+_1 \rightarrow 2^+_1$ & 28.9 & 15 \\
$6^+_1 \rightarrow 4^+_1$ & 18.8 & 6.9 \\
$2^+_1 \rightarrow 2^+_2$ & 3.2 & 13.3 \\
$4^+_1 \rightarrow 2^+_2$ & 4 & 1.6 \\
$2^+_3 \rightarrow 0^+_1$ & 1.6 & 3 \\
\hline
\end{tabular}
\end{center}
\end{table}

\vskip 2cm

\begin{table}
\begin{center}

\caption{Deformed Shell Model results for  half-lives $T_{1/2}^{2\nu}$ and
the corresponding phase space factors $G_{2\nu }$ for possible decay modes
for $^{84}$Sr $\rightarrow  ^{84}$Kr. Note that the natural abundance of
$^{84}$Sr is 0.56\% \cite{Boh05}.  The $G_{2\nu }$ and half-lives are
calculated using $g_A/g_V=1$. For half-lives given in the brackets see
text.}

\begin{tabular}{llc}
\hline 
Mode & $G_{2\nu }$ (in yr$^{-1}$) & $T_{1/2}^{2\nu}$ (in yr) from DSM \\
\hline
$\beta ^{+}$EC & $\;\;\;\;\;1.504 \times 10^{-24}$ &  $\;\;\;\;1.2 \times 
10^{26}$ ($5.41 \times 10^{25}$) \\ 
ECEC & $\;\;\;\;\;4.367 \times 10^{-23}$ & $\;\;\;\;4.16\times 10^{24}$ 
($1.87\times 10^{24}$) \\
\hline
\end{tabular}
\end{center}
\end{table}

\newpage

\begin{figure*}[ht]
\includegraphics[width=6in]{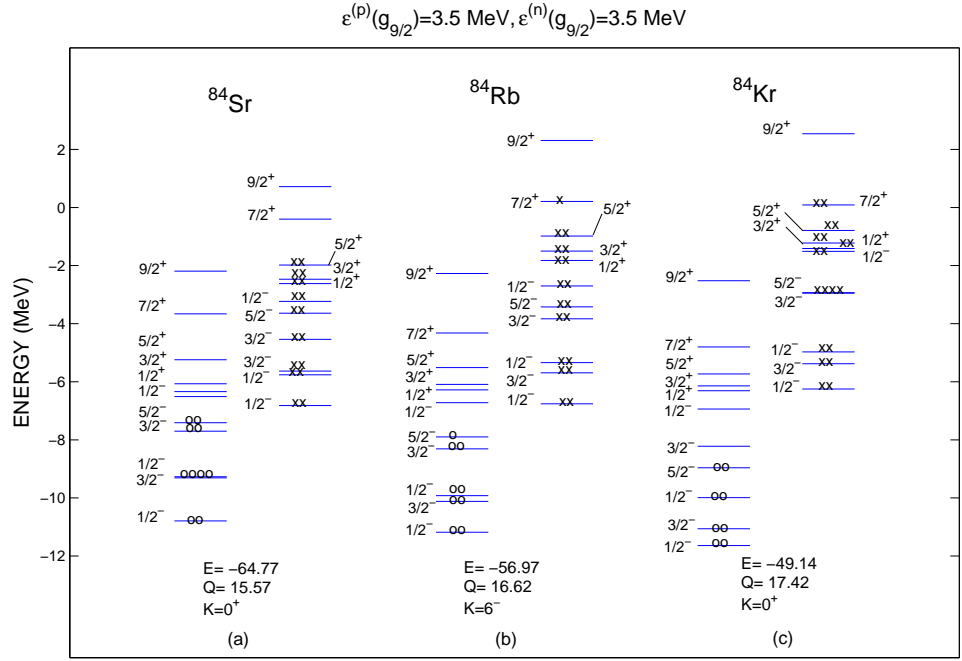}

\caption{HF single particle spectra for $^{84}$Sr, $^{84}$Rb and $^{84}$Kr
for $\epsilon^{(p)}(^1g_{9/2}) = \epsilon^{(n)}(^1g_{9/2}) =3.5$ MeV.  In the
figures circles represent protons and crosses represent neutrons. The
Hartree-Fock energy  ($E$) in MeV, mass quadrupole moment ($Q$) in units of
the square of the oscillator length parameter and the total $K$ quantum
number of the lowest intrinsic states are given in the figure.}

\label{fig1}
\end{figure*}

\newpage

\begin{figure*}[ht]

\includegraphics[width=6in]{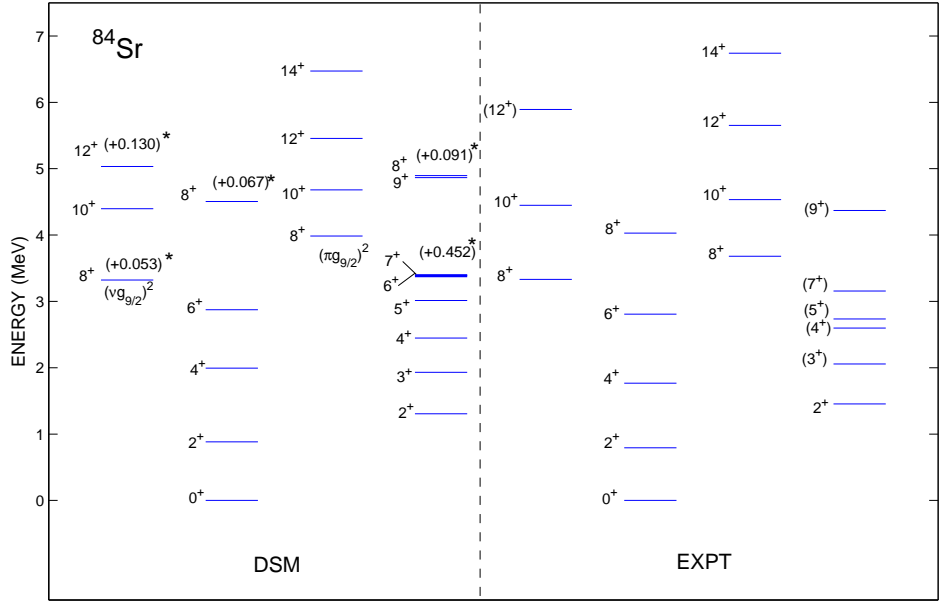}

\caption{(a) Calculated and (b) experimental spectra for $^{84}$Sr. 
Experimental data are from \cite{dew-82,nndc}. Note that the DSM spectrum is
obtained using $[\epsilon^{(p)}(^1g_{9/2}),\epsilon^{(n)}(^1g_{9/2})] 
=[3.5, 1.5]$MeV.  The numbers in the brackets (over a few levels)  indicate
the difference between the calculated energies and those obtained with the
choice $[\epsilon^{(p)}(^1g_{9/2}),\epsilon^{(n)}(^1g_{9/2})]  =[3.5,
3.5]$MeV. See text for details.}

\label{fig2}
\end{figure*}

\newpage

\begin{figure*}[ht]
\includegraphics[width=6in]{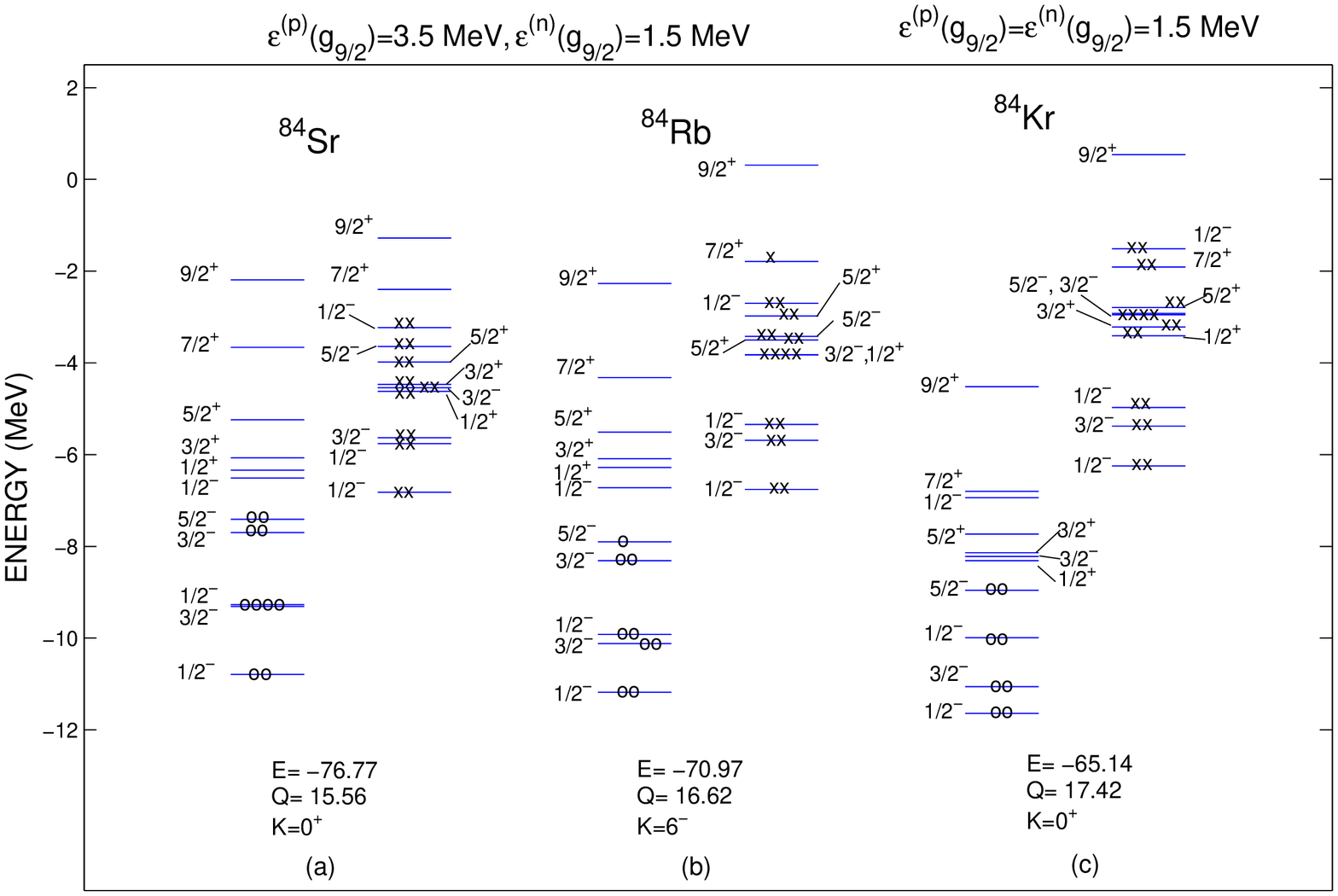}

\caption{HF single particle spectra for $^{84}$Sr, $^{84}$Rb and $^{84}$Kr.
Note that $[\epsilon^{(p)}(^1g_{9/2}),\epsilon^{(n)}(^1g_{9/2})]  =[3.5,
1.5]$MeV for $^{84}$Sr and $^{84}$Rb nuclei and 
$[\epsilon^{(p)}(^1g_{9/2}),\epsilon^{(n)}(^1g_{9/2})]  =[1.5, 1.5]$MeV for
$^{84}$Kr. In the figures circles represent protons and crosses represent
neutrons. The Hartree-Fock energy  ($E$) in MeV, mass quadrupole moment
($Q$) in units of the square of the oscillator length parameter and the
total $K$ quantum number of the lowest intrinsic states are given in the
figure. See text for details.}

\label{fig3}
\end{figure*}

\newpage

\begin{figure*}[ht]
\includegraphics[width=6in]{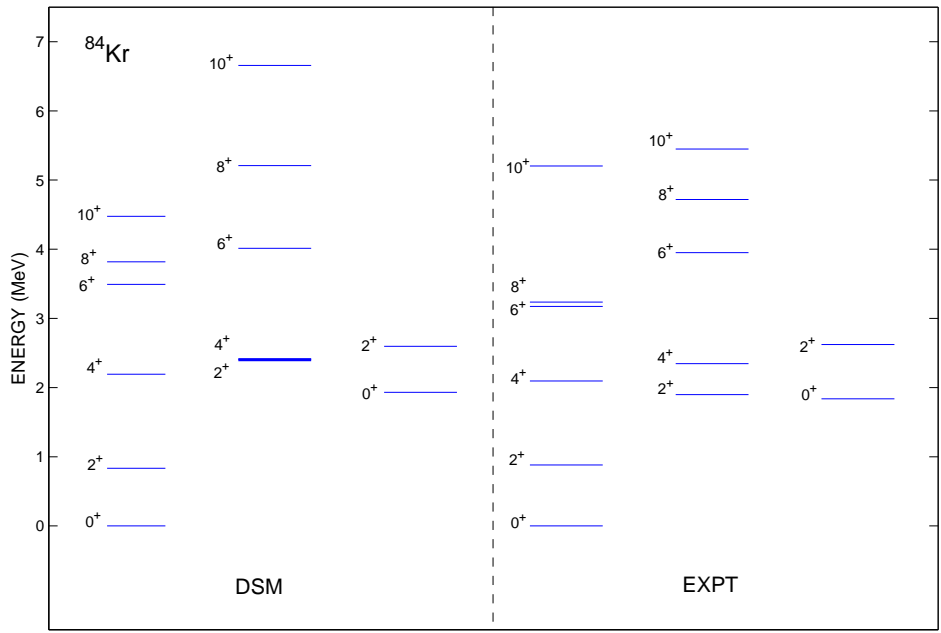}

\caption{(a) Calculated and (b) experimental spectra for $^{84}$Kr. The
experimental data are from \cite{nndc}. See text for details.}

\label{fig4}
\end{figure*}

\ed